\documentclass[a4paper,11pt]{article}
\usepackage{jheppub} 

\usepackage[T1]{fontenc}
\usepackage{amsthm}
\usepackage{amsmath}
\newtheorem{theorem}{Theorem}
\usepackage{enumitem}
\usepackage{comment}
\title{\boldmath Celestial gluon and graviton OPE at loop level }

\author[]{Hare Krishna}

\affiliation[]{C. N. Yang Institute for Theoretical Physics, Stony Brook University,\\ Stony Brook, NY 11794, USA}

\emailAdd{harekrishna.harekrishna@stonybrook.edu,}

\newcommand{\f}{\frac}

\newcommand{\non}{\nonumber}
\newcommand{\be}{\begin{eqnarray}}
\newcommand{\ee}{\end{eqnarray}}

\abstract{In this paper, we analyze the loop corrections to celestial OPE for gluons and gravitons. Even at the loop level, the soft gluons and gravitons have conformal dimensions $\Delta=1-\mathbb Z_{\geq 0}$. The only novelty is the presence of higher poles. At one loop level, there are two types of conformal soft gluons with a single pole and a double pole in the $\Delta$ plane. The celestial OPEs are obtained using the collinear splitting functions. In the case of gluons, the splitting functions receive loop corrections. After taking the holomorphic soft limit, we find the OPE of conformal soft gluons. We find a novel mixing of simple and double poles soft gluon operators in the OPE. In the case of gravitons, where splitting functions are known to be all loop exact, we still find a wedge algebra of $w_{\infty}$ which is in addition to the wedge algebra of $w_{1+\infty}$ already found by Strominger.   }

\begin{document} 
\hfill       YITP-SB-2023-31

\maketitle
\flushbottom

\section{Introduction}

 The quantum gravity in asymptotically AdS spacetime can be described by a conformal field theory living on its conformal boundary. There has been a lot of interest in generalizing these ideas to asymptotically flat spacetime (some of the developments have been made in \cite{Atanasov:2021cje,Ball:2021tmb,Crawley:2021ivb,Donnay:2020guq,Guevara:2019ypd,Jiang:2021ovh,Pasterski:2016qvg,Pasterski:2017kqt,Pasterski:2017ylz,Pasterski:2021raf,Pasterski:2021rjz,Banerjee:2022wht,Pasterski:2022djr,Pate:2019lpp,Pate:2019mfs,Raclariu:2021zjz} and \cite{Arkani-Hamed:2020gyp,Himwich:2020rro,Ball:2019atb,Pano:2023slc,Mcloughlin:2022ljp,Travaglini:2022uwo,Donnay:2022sdg,Pasterski:2021fjn,Pasterski:2020pdk,Casali:2020vuy}). In 4d asymptotically flat spacetime, the "CFT" (called CCFT) is proposed to live on the celestial sphere $S^2$. The development in this direction is mainly bottom-up. The scattering in bulk 4d spacetime can be written as a correlation function of operators on the celestial sphere. The operators on the celestial sphere are defined by extrapolating the bulk fields. For massless fields, the operators on the celestial sphere are written in terms of a Mellin variable conjugate to the energy. For massive fields, one uses the bulk to boundary propagator of AdS$_3$ to go from momentum basis to celestial basis. The celestial basis diagonalizes boost rather than momenta.  The operator product expansion of these operators can be obtained by the collinear limit of the amplitude where two particles become parallel. Using the existing results in the collinear limit of amplitudes in gauge theory and gravity, these authors \cite{Pate:2019lpp,Guevara:2021abz,Strominger:2021lvk,Fotopoulos:2019vac,Fan:2019emx} found OPE coefficients. Most of the results in the literature are based on tree-level scattering. The complete description of the CCFT which is dual to QCD or gravity has not been constructed. Nevertheless, there are some interesting toy models for CCFT mainly in the self-dual sector of QCD and gravity \cite{Adamo:2021lrv,Costello:2022jpg,Costello:2023hmi} and even on the string world-sheet \cite{Krishna:2023ecp}. \\
 
 In the exploration of CCFT, soft theorem plays an important role. The statement of soft theorems becomes the Ward identity on the celestial sphere \cite{Strominger:2013jfa,Strominger:2013lka,Lysov:2014csa,Kapec:2014opa,Kapec:2016jld,Pasterski:2015tva,Kapec:2014zla,Kapec:2015vwa,He:2015zea,Kapec:2015ena,Strominger:2015bla,Dumitrescu:2015fej,He:2017fsb,Strominger:2017zoo,Nande:2017dba,Kapec:2017tkm,Pate:2017vwa}. Hence, these theorems manifest some conserved currents in CCFT. The scattering states in flat space-time are usually written in momentum space. The massless states are written as $|p^{\mu},\pm h\rangle$, here $p^{\mu}$ is the momentum and $\pm h$ is the helicity of the particle. This basis diagonalizes the translation generator. For CCFT, it is useful to look at the states which diagonalize Lorentz transformation. The reason is the Lorentz group $SO^+(3,1)\cong SL(2,C)/Z_2$ becomes the global conformal transformation on the celestial sphere. So, the states in CCFT are boost eigenstates rather than momentum eigenstates. One can go from a momentum eigenbasis to a boost eigenbasis by doing a Mellin transformation. Pasterski and Shao showed that boost eigenstates form a complete basis with Klein-Gordon norm if $\Delta \in 1+ i \lambda, \,  \lambda \in \mathbb{R}$ \cite{Pasterski:2017kqt}. There is a special place for soft operators/theorems.\\
 
 The study of soft theorems began in the early work of \cite{PhysRev.140.B516}. And there has been a lot of renewed interest since the work of Cachazo and Strominger \cite{Cachazo:2014fwa,Lysov:2014csa,Kapec:2014opa,He:2014laa}. Soft theorems have been generalized to the generic theory of quantum gravity (see \cite{AtulBhatkar:2018kfi,Chakrabarti:2017ltl,Chakrabarti:2017zmh,Laddha:2017ygw,Laddha:2018rle,Sahoo:2018lxl,Sen:2017xjn,Sen:2017nim,Sen:1981sd,Sahoo:2021ctw,Sahoo:2020ryf}). One can also write the conformal soft theorem using the Mellin transformation of momentum space soft theorem. The conformal soft operators have $\Delta=\{1,0,-1,-2,\cdots\}$  \cite{Donnay:2018neh,Himwich:2019dug,Pate:2019lpp,Puhm:2019zbl,Donnay:2020guq,Pasterski:2021dqe,Pano:2021ewd}. This is true even if we incorporate the loop corrections in the scattering amplitude. So, these soft operators are genuine currents in CCFT.  Recently, it has been shown that only a discrete set of states with $\Delta \in \mathbb{Z}$ (conformal primary wave functions) form a basis of massless states \cite{Cotler:2023qwh,Freidel:2022skz} under some assumptions on the fall-off behavior of these primary wave-functions. Equipped with the knowledge of operator content, the next step is to understand the OPE of these operators. The problem of finding the OPE translates to the collinear limit of the scattering amplitudes. In the collinear limit, the amplitudes factorize in terms of the lower point amplitudes combined with splitting functions. The collinear splitting function (leading) for graviton scattering in Einstein gravity is all-loop exact. In QCD, the collinear splitting functions receive loop corrections. One goal of this paper is to understand the implication of these corrections on the operator product expansion of conformal soft operators in gauge theory and gravity.  See some previous works on loop correction to celestial amplitudes in refs. \cite{Donnay:2023kvm,Gonzalez:2020tpi,Ball:2021tmb}  \\

In this paper, we are going to study the 1-loop correction to the OPE of gluons and gravitons on the celestial sphere. Bhardwaj et al.\cite{Bhardwaj:2022anh} have found the loop correction to the celestial gluon OPE which has generic conformal dimensions. We will take the soft limit of these OPEs. We have two types of conformal soft operators with simple poles and double poles in the $\Delta$ plane. Both operators have the same conformal weight. We will find the OPE of these conformal soft operators. At the loop level, these OPEs will become more complicated, and their residues are written in terms of Polylogs and Harmonic numbers. In the graviton OPE, the tree level OPE is all loop exact because the splitting functions don't receive any loop correction. But soft operators do receive loop correction and we have two sets of conformal soft graviton operators with simple poles and double poles. We found the OPE of these conformal soft operators and their mode algebras. This gives us another set of wedge subalgebra of $w_{\infty}$ algebra on top of $w_{1+\infty}$ algebra already found by Strominger \cite{Strominger:2021lvk}. \\

In section \ref{soft operator}, we will provide details about the soft operators at the loop level. The conformal soft operator still has the discrete conformal dimension but the simple pole in OPE will be generalized to higher poles. This is a huge simplification compared to generic hard operators, which lie on the principal series. In section \ref{loop ope}, we will review the OPE of gluons with generic conformal dimensions which was obtained in \cite{Bhardwaj:2022anh}. In section \ref{section4}, we will take the soft limit of these OPEs. There are three sets of OPE to consider, which are discussed in subsections. In section \ref{section6}, we studied the conformal soft graviton OPE and found a new set of $w_{\infty}$ algebra. We finish the article with discussions and outlook in section \ref{outlook}. In appendix \ref{Mellin requirement}, we have discussed the properties of Mellin and inverse transform which are essential to our discussion. The details regarding the soft gluon theorem at one loop are sketched in appendix \ref{soft gluon}. 

\textbf{Note added:-} We are aware of an upcoming work by Bhardwaj and Akshay Srikant on "Loop corrections, soft factors, and logarithmic descendants in celestial CFT" which may have overlapping results to ours.

\section{Soft operators at loop level}
\label{soft operator}
The amplitudes involving massless particles in $D=4$ spacetime have infrared divergences. 
The first question we need to address is the meaning of soft theorems if amplitudes are divergent. There are many ways to address this question. Bern et al. \cite{Bern:2014oka} started with the divergent amplitudes and have divergent soft factors. These soft factors upon Fourier transform correspond to memory effects on the asymptotic infinity \cite{Strominger:2017zoo,Pasterski:2015tva,Strominger:2014pwa}. These memory effects can be observed experimentally in future experiments. Therefore, an infrared finite observable is required for this purpose. For an infrared-finite soft factor, we follow a strategy laid out in the very early work of Grammer and Yennie \cite{PhysRevD.8.4332}. It was proven to be very useful in QCD in understanding the Sudakov form factors among other things \cite{Sen:1981sd}. These methods have been revived to understand the soft theorem \cite{Sahoo:2018lxl,Krishna:2023fxg}. The goal of Grammer and Yennie was to find a good prescription to compute the IR finite S matrix. The concept entails expressing the massless gauge boson propagator in two distinct parts, denoted as $K$ and $G$. The amplitudes with only $G$ photon/graviton propagator are infrared finite. The amplitudes with $K$ propagators contain all the divergences and get exponentiated. Now, the idea is to factor out these diverging exponentials and define a finite $S$- matrix. The soft theorem relates two sets of amplitudes with and without extra soft particles denoted as $\mathcal{A}_{n+1}$ and $\mathcal{A}_n$. For example, in QED this factorization can be written as
\begin{eqnarray}
\mathcal{A}^{(N)}=\exp\lbrace \mathrm{ K_{em}}\rbrace\ \mathcal{A}^{(N)}_{\text{IR-finite}}\qquad \qquad  \mathcal{A}^{(N+1)}=\exp\lbrace \mathrm{K_{em}}\rbrace\ \mathcal{A}^{(N+1)}_{\text{IR-finite}}\ . 
\label{KG decomposition}    
\end{eqnarray}
In the above equation, $\mathrm{K_{em}}$ contains the full IR divergent contributions. It is important to note that we have the same $\mathrm{K_{em}}$ factor in both amplitudes. The explicit expression of
 $\mathrm{K_{em}}$ 
\begin{eqnarray}
 \mathrm{K_{em}}=  \ \f{i}{2}\sum_{i=1}^{N}\  \sum_{\substack{j=1\\j\neq i}}^{N}e_{i}e_{j}\ \int \f{d^{4}\ell}{(2\pi)^{4}} \ \f{1}{\ell^{2}-i\epsilon}\ {(2p_i-\ell)\cdot(2p_j+\ell) \over (2p_i\cdot\ell -\ell^2+i\epsilon) 
(2p_j\cdot \ell+\ell^2-i\epsilon)}
\label{K_em}\ .   
\end{eqnarray}
Here $e_i$ and $p_i$ are the electric charges and momenta of the particle.
After this factorization, we can compare the IR finite amplitudes as
\begin{eqnarray}
 \mathcal{A}^{(N+1)}_{\text{IR-finite}}= \text {Soft factor} \times \mathcal{A}^{(N)}_{\text{IR-finite}}   
\end{eqnarray}
One can study these soft factors at various orders in QFTs as well as in gravity. The leading, subleading and subleading soft factors at various loops have been analysed in \cite{Sahoo:2021ctw,Sahoo:2018lxl,Krishna:2023fxg,Ghosh:2021bam,AtulBhatkar:2018kfi,AtulBhatkar:2019vcb,AtulBhatkar:2020hqz}. In QCD, the Grammer-Yennie (KG) decomposition was studied a long time ago by Sterman and Collins \cite{Sterman:1981jc,Collins:1981ta} and then used by Sen to study the form factors \cite{Sen:1981sd}. We use these methods to learn more about the soft theorems. 

In this article, we will analyze one-loop corrections to the soft theorem. For gluons, it was established by Catani \cite{Catani:2000pi,Bern:1998sv,Bern:2014oka} that even the leading soft gluon current gets loop corrected. The tree level contribution to soft factor is proportional to $g_{YM}$ while one loop part comes with $g^3_{YM}$. They have the explicit IR divergence in the soft factor because they were analyzing the theory in dimensional regularization. In the appendix \ref{soft gluon}, we will analyze the gluon soft theorem directly in $D=4$ using the Grammar-Yennie decomposition. After factoring out the divergent parts, the soft theorem can be written as
\begin{eqnarray}
    \mathcal{A}_{N+1}^{\mathrm{1-loop,\,IR-finite}} =\Bigg(S^{\mathrm{tree}}_0 + S^{\mathrm{tree}}_1\Bigg) \mathcal{A}_{N}^{\mathrm{1-loop,\,IR finite}}+\Bigg(S^{\mathrm{1-loop}}_0 + S^{\mathrm{1-loop}}_1\Bigg) \mathcal{A}_{N}^{\mathrm{tree}}
\end{eqnarray}
Here $S_0$ and $S_1$ represent the leading and subleading soft gluon factor. The only fact that we are going to need is $S_0^{\mathrm{tree}} \propto \frac{1}{\omega}$ and $S_0^{\mathrm{1-loop}} \propto \frac{1}{\omega}\mathrm{Log}[\omega]$. It is reasonable to think that  $S_1^{\mathrm{tree}} \propto \omega^0$ and $S_1^{\mathrm{1-loop}} \propto \mathrm{Log}[\omega]$. Here $\omega$ is the energy of the extra gluon in the $n+1$ point amplitude.\\

This concludes our discussion of soft factors in the momentum space. But for celestial CFT, we need to analyze the soft theorems in boost eigenstates. This can be achieved by doing the Mellin transform of all the incoming/outgoing states.
\begin{eqnarray}
\label{mellin 1}
    \mathcal{O}_{\Delta}\equiv\int_0^{\infty} \frac{d \omega}{\omega} \omega^{\Delta} f(\omega)
\end{eqnarray}
Here $f(\omega)$ could be a plane wave or a wave packet. The soft operators are special in celestial holography as they have $\Delta \in \{1,0,-1,-2,\cdots\}$. Usually, the soft operators are identified by doing the Mellin transformation of whole amplitudes. In the soft limit, amplitudes have a formal asymptotic expansion (we know only up to a few orders) in $\omega^i \mathrm{Log}^j[\omega]$. Doing the Mellin transformation of soft factors produces poles in $\Delta$ plane (see appendix \ref{Mellin requirement} for details about loop corrections). In this section, we will truncate the expansion at one loop. Then, the soft operator has the following expansion
\begin{eqnarray}
      O^{\Delta,a}(z,\bar{z})\sim  c_{i,0}\frac{O_{-1}^{i,a}(z,\bar{z})}{(\Delta+i)}+c_{i,1}\frac{O_{-2}^{i,a}(z,\bar{z})}{(\Delta+i)^2}
  \end{eqnarray} 
Here the various labels of $O_{-1}^{i,a}(z,\bar{z})$ are as follows. The $i$ label is related to the dimension of the conformal soft operator, $a$ labels the color index, and an additional index which is $-1$ describes the order of the poles. The $c_{i,0}$ and $c_{i,1}$ are constants and could be a function of coupling constants as soft factors depend on coupling constants. Just as an example $i=2$ means the $\mathrm{(sub)^3 }$ leading theorem which corresponds to $\Delta=-2$. \\

But for gravitons, the leading soft theorem is all loop exact. Hence, the leading soft operator will have only simple poles in the Mellin basis. But subleading and higher soft theorem do receive 1-loop correction. After truncating the expansion at one loop, the soft graviton operator has the following expansion 
\begin{eqnarray}
      G^{\Delta}(z,\bar{z})\sim  c_{i,0}\frac{G_{-1}^{i}(z,\bar{z})}{(\Delta+i)}+c_{i,1}\frac{G_{-2}^{i}(z,\bar{z})}{(\Delta+i)^2}
  \end{eqnarray}
  But here $i \in \{0,-1,-2,\cdots\}$. The conformal soft operators are defined as the residues at the poles. The coefficients $G_{-1}^{i}(z,\bar{z})$ and $G_{-2}^{i}(z,\bar{z})$ will become the conformal graviton operator. The presence of higher poles was already anticipated in prior work of \cite{Donnay:2022sdg,Guevara:2019ypd}. However, the connection to the explicit conformal operator was not established. In this work, we have identified the conformal operator which corresponds to simple and higher poles. Having identified the operators on the celestial sphere, now we will try to find the OPE of these operators. This can be achieved by finding the collinear limit of the amplitudes.

\section{Loop level gluon OPE using Collinear limit}
\label{loop ope}
 In this section, we will study the OPE of the positive helicity gluons.\footnote{ The negative helicity particles generate the analogous currents to positive helicity ones. The OPE of opposite helicity particles depends on the order of limit.} The Lorentz symmetry of Minkowski spacetime manifests as global conformal transformation $SL(2,R)_L\times SL(2,R)_R$ on the celestial sphere. The $SL(2,R)_L$ will get enhanced to Virasoro thanks to the subleading soft gravitons which generate superrotations. But we only have the chiral half $Vir_L$ as we are working with just positive helicity.\footnote{ $Vir_R$ are generated by negative helicity gravitons/gluons. We will treat $z$ and $\bar{z}$ as independent variables, hence the right signature of spacetime will be $(2,2)$. It is also reasonable to think that these symmetries will survive the loop correction. The reason is  $SL(2,R)_R$ is part of the Lorentz group and $Vir_L$ is generated by a loop-level subleading soft theorem which is universal and one loop exact.} Hence, we will work in  $Vir_L\times SL(2,R)_R $ formalism. The soft currents form a representation of these algebras.

We will start with positive helicity gluons with conformal dimension $\Delta=k$ which transforms under $(2-k)$ dimensional representation of $SL(2,R)_R$ \cite{Guevara:2021abz,Strominger:2021lvk}. The states with lowest weight has $\bar{h}=\frac{k-1}{2}$ and the states with highest weight has $\bar{h}=\frac{1-k}{2}$. The mode expansion of these currents/operators can be written as
\begin{equation}
    O^{a,+}_k(z,\bar{z})=\sum_{n=\frac{k-1}{2}}^{\frac{1-k}{2}}\frac{O^{a,+}_{k,n}(z)}{\bar{z}^{n+\frac{k-1}{2}}}
\end{equation}
here $\Delta=k=1,0,-1,\cdots$. We are keeping the arbitrary $z$ dependence of these operators. Now we define conformal soft operators as the limits of the soft operators. 
\begin{equation}
    R_{k,n}^{a,+}(z):= \lim_{\epsilon \rightarrow 0} \epsilon \,\, O^{a,+}_{k+\epsilon,n}(z), \quad k=0,1,-1,-2..., \quad \frac{k-1}{2} \leq n\leq \frac{1-k}{2}
\end{equation}

More precisely the conformal soft operators are defined as the residues at the poles. We have added one more label to highlight the simple and double poles.
\begin{eqnarray}
 R^{a,+}_{k,n,-1}(z)    =\oint_k \frac{d \Delta}{2 \pi i} O^{a,+}_{\Delta,n}(z) 
\end{eqnarray}
At the loop level, we have higher poles; therefore, the definition would be
\begin{eqnarray}
  R^{a,+}_{k,n,-2}(z)   =\oint_k \frac{d \Delta}{2 \pi i} (\Delta-k)O^{a,+}_{\Delta,n}(z) 
\end{eqnarray}
The corresponding conformal soft operators after summing over the modes can be written as
\begin{equation}
\label{conformally soft}
    R^{a,+}_{k,-1}(z,\bar{z})=\sum_{n=\frac{k-1}{2}}^{\frac{1-k}{2}} \frac{ R^{a,+}_{k,n,-1}(z)}{\bar{z}^{n+\frac{k-1}{2}}}, \quad 
     R^{a,+}_{k,-2}(z,\bar{z})=\sum_{n=\frac{k-1}{2}}^{\frac{1-k}{2}} \frac{ R^{a,+}_{k,n,-2}(z)}{\bar{z}^{n+\frac{k-1}{2}}} 
\end{equation}
Both operators $R^{a,+}_{k,-1}(z,\bar{z})$ and $R^{a,+}_{k,-2}(z,\bar{z})$ (Just to be clear the labels here are as follows- $a$ is the color index, $+$ is the helicity of the gluons, $k$ is the conformal dimension $\Delta$ of the conformal gluons and $-1$ and $-2$ labels the order of the poles) have conformal weight as 
    $(h,\bar{h})=\Bigg(\frac{k+1}{2},\frac{k-1}{2}\Bigg)$. 
\subsection*{Gluon OPE}
We will start with the OPE for positive helicity gluons as calculated in \cite{Bhardwaj:2022anh} (see eq  3.52) \footnote{The addition of $SL(2,R)_R$ descendants to \eqref{ope gluons positive both raw} is very subtle at loop level. We thank Rishabh Bhardwaj for the discussion on this point.}
\begin{eqnarray}
\label{ope gluons positive both raw}
    O^{a,+}_{\Delta_1}(z_1,\bar{z}_1) O^{b,+}_{\Delta_2}(z_2,\bar{z}_2)&&\sim i \frac{f^{abc}}{z_{12}} \Bigg[1+ a H_{0,+}^{(1)}+a C_{2,+}^{(1)} \hat{\mathcal{D}}_{12}^2\Bigg]  C_+^{(0)} O^{c,+}_{\Delta_1+\Delta_2-1} \nonumber\\
    &&+i \frac{f^{abc}}{z_{12}} a C_{0,-}^{(1)} O^{c,+}_{\Delta_1+\Delta_2-1} +i f^{abc} \frac{a}{2}\frac{\bar{z}_{12}}{z_{12}^2} C_{0,-}^{(1)} O^{c,-}_{\Delta_1+\Delta_2-1}.
\end{eqnarray}
The last term in the above OPE is a little different, as it contains the $-$ helicity, which is unexpected based on the understanding of tree-level OPE. The terms in the OPE written above are the most singular in ($z_{12}$). One could ask whether this OPE has a convergent expansion in $z_{12}$. We don't know the answer to this question.  All the factors in the above OPE \eqref{ope gluons positive both raw} are listed here
\begin{eqnarray}
  && C_+^{(0)}=B(\Delta_1-1,\Delta_2-1),\quad  a= \frac{g^2 N_c}{8 \pi^2}, \quad C_{2,+}^{(1)}=-2,\nonumber\\
  &&  H_{0,+}^{(1)}=-\frac{\pi^2}{12}+2 \partial_{\Delta_1}\partial_{\Delta_2}, \qquad  C_{0,-}^{(1)}=\frac{1}{3}(1+\frac{n_s}{N_c}-\frac{n_f}{N_c}) B(\Delta_1,\Delta_2)\non\\
&&\hat{\mathcal{D}}_{12}=\partial_{\Delta_1}+\partial_{\Delta_2}+\partial_{\Delta}+\frac{1}{2} \mathrm{Log}[- z_{12}\bar{z}_{12} \frac{\Lambda^2}{\mu^2}]
\end{eqnarray}
The operator $\hat{\mathcal{D}}_{12}$ has a derivative $\partial_{\Delta}$ which acts on $O^{c,+}_{\Delta}$. In this case the operator $O^{c,+}_{\Delta}$ is assumed to be independent of $\Delta_1,\Delta_2$. After taking the derivative we should evaluate the expression at $\Delta=\Delta_1+\Delta_2-1$. The $\partial_{\Delta_1},\partial_{\Delta_2}$ derivative acts only on the $B(\Delta_1-1,\Delta_2-1)$. In the above equation, $n_s$ and $n_f$ correspond to the number of scalars and fermions in the fundamental representation of the gauge group.  \\

\textbf{Some comments about the Log term}
:-The appearance of $\mathrm{log}$ can be traced back to the loop-level splitting functions. Splitting functions have two kinds of terms: factorizing and non-factorizing. In the OPE, the Log term came from the non-factorizing part. The $\mu$ represents the renormalization scale. The parameter $\Lambda$ is introduced to make $\omega$ is dimensionless.\\

From the OPE \eqref{ope gluons positive both raw}, We can see that OPE coefficients have a pole whenever $\Delta_1,\Delta_2=\{1,0,-1,\cdots\}$. The conformal soft operators are defined by the residue at these poles.
\section{Conformal gluons OPE}
\label{section4}
In this section, we are going to study the OPE of conformally soft gluons which are defined as the residues at the poles. There are two types of residues: for the simple pole, we have $R^{a,+}{k,-1}(z,\bar{z})$, and for double poles, we have $R^{a,+}{k,-2}(z,\bar{z})$. 
With these conformal soft operators, we encounter three different types of OPE
\begin{itemize}
    \item $R^{a,+}_{k,-1}(z_1,\bar{z}_1)R^{b,+}_{l,-1}(z_2,\bar{z}_2)$
    \item $R^{a,+}_{k,-2}(z_1,\bar{z}_1)R^{b,+}_{l,-2}(z_2,\bar{z}_2)$
    \item  $R^{a,+}_{k,-2}(z_1,\bar{z}_1)R^{b,+}_{l,-1}(z_2,\bar{z}_2)$
\end{itemize}
We will study these OPEs in the subsequent subsections.
\subsection{First type OPE}
First, we are going to analyse the OPE of $R^{a,+}_{k,-1}(z_1,\bar{z}_1)R^{b,+}_{l,-1}(z_2,\bar{z}_2)$. This is the OPE of conformal soft operators having simple poles. This can be evaluated from the OPE in \eqref{ope gluons positive both raw} using the contour integration. The contour integral that we are taking is $\oint_{(k} \oint_{l)}=\frac{1}{2}(\oint_k\oint_l+\oint_l \oint_k)$. This way of taking the contour integral matches with the direct computation by making the dimension as $k+\epsilon,l+\epsilon$ and taking the $\epsilon \rightarrow 0$ limit whenever this limit is non-vanishing.\\
Let's start with the \textbf{tree} type term in the \eqref{ope gluons positive both raw} \footnote{Here by tree type, we mean the term without any derivative with respect to $\Delta,\Delta_1,\Delta_2$}  
\begin{align}
 \label{firstzeroder}
    & R^{a,+}_{k,-1}(z_1,\bar{z}_1)R^{b,+}_{l,-1}(z_2,\bar{z}_2) \sim  \oint_{(k} \oint_{l)}O^{a,+}_{\Delta_1}(z_1,\bar{z}_1)  O^{b,+}_{\Delta_2}(z_2,\bar{z}_2)\non\\
    & \sim i \frac{f^{abc}}{z_{12}}\frac{(2-k-l)!}{(1-k)!(1-l)!} R_{k+l-1,-1}^{c,+}(z_2,\bar{z}_2)\non\\
 &+ i \frac{g^2 N_c}{8 \pi^2}\frac{f^{abc}}{z_{12}}  
\Bigg[\frac{1}{3}(1+\frac{n_s}{N_c}-\frac{n_f}{N_c})\frac{(-k-l)!}{(-k)!(-l)!}-\frac{\pi^2}{12}\frac{(2-k-l)!}{(1-k)!(1-l)!}\Bigg]  R_{k+l-1,-1}^{c,+}\non\\
&-i \frac{g^2 N_c}{16 \pi^2}\frac{(2-k-l)!}{(1-k)!(1-l)!}\frac{f^{abc}}{z_{12}}  
\Bigg[\mathrm{Log}^2[- z_{12}\bar{z}_{12} \frac{\Lambda^2}{\mu^2}]\Bigg]  R_{k+l-1,-1}^{c,+}\non\\
&+i \frac{g^2 N_c}{16 \pi^2}\frac{f^{abc} }{z_{12}^2}  \frac{1}{3}(1+\frac{n_s}{N_c}-\frac{n_f}{N_c})  \frac{(-k-l)!}{(-k)!(-l)! }\bar{z}_{12}  R_{k+l-1,-1}^{c,-}  
\end{align}


Next, we will consider the terms in \eqref{ope gluons positive both raw} which contain \textbf{two derivatives} (in total) which might be $\partial_{\Delta_1}\partial_{\Delta_2}$ acting on the Beta function or $\partial_{\Delta}\partial_{\Delta}$ acting on the operator or any other combinations. All the contributions are summed and written compactly as
\begin{align}
 \label{firsttwoder}
 & R^{a,+}_{k,-1}(z_1,\bar{z}_1)R^{b,+}_{l,-1}(z_2,\bar{z}_2)\sim\oint_{(k} \oint_{l)} O^{a,+}_{\Delta_1}(z_1,\bar{z}_1)  O^{b,+}_{\Delta_2}(z_2,\bar{z}_2)\supset  i \frac{(-g^2 N_c)}{\pi^2}\frac{f^{abc}}{z_{12}}   
   \frac{\Gamma (-k-l+3) }{ \Gamma (2-l) \Gamma (-k+2)} \nonumber\\
    &\Bigg(-\psi ^{(0)}(-k+2)^2 \Big[R_{k+l-1,-1}^{c,+}+3 R_{k+l-1,-2}^{c,+} \psi ^{(0)}(-k-l+3)\Big]-R_{k+l-1,-1}^{c,+} \left(H_{1-l}-H_{-k-l+2}\right){}^2 \non\\
     &
   +2 R_{k+l-1,-1}^{c,+} \psi ^{(0)}(-k+2) \psi ^{(0)}(-k-l+3)-R_{k+l-1,-1}^{c,+} \psi
   ^{(0)}(-k-l+3)^2-2 R_{k+l-1,-1}^{c,+} \psi ^{(1)}(-k-l+3)\non\\
   &+R_{k+l-1,-1}^{c,+} \psi ^{(1)}(-k+2)+R_{k+l-1,-1}^{c,+} \psi
   ^{(1)}(2-l)+R_{k+l-1,-2}^{c,+} \left(H_{1-l}-H_{-k-l+2}\right) \non\\
   &\Big[\left(H_{1-l}-H_{-k-l+2}\right){}^2+3
   \psi ^{(1)}(-k-l+3)-3 \psi ^{(1)}(2-l)\Big]+3 R_{k+l-1,-2}^{c,+} \psi ^{(0)}(-k+2) \psi ^{(0)}(-k-l+3)^2\non\\
   &+3 R_{k+l-1,-2}^{c,+}
   \psi ^{(0)}(-k+2) \Big[\psi ^{(1)}(-k-l+3)-\psi ^{(1)}(-k+2)\Big]-R_{k+l-1,-2}^{c,+} \psi ^{(0)}(-k-l+3)^3 \psi
   ^{(0)}(-k-l+3)\non\\
   &+3 R_{k+l-1,-2}^{c,+}\Big[\psi ^{(1)}(-k+2)-\psi ^{(1)}(-k-l+3)\Big]-2 R_{k+l-1,-2}^{c,+} \psi ^{(2)}(-k-l+3)\non\\
   &+R_{k+l-1,-2}^{c,+} \psi
   ^{(0)}(-k+2)^3+R_{k+l-1,-2}^{c,+} \psi ^{(2)}(-k+2)+R_{k+l-1,-2}^{c,+} \psi ^{(2)}(2-l)\Bigg)\non\\    
\end{align}

 The OPE of conformal operators having simple poles generates the conformal operator having double poles. This is the curious aspect of our calculation. The $\psi ^{(i)}(x)$ are the poly gamma functions and $H_{x}$ are the Harmonic numbers. These functions can be written as
\begin{align}
    \psi(z)=\frac{\Gamma'(z)}{\Gamma(z)}, \quad \psi^{(n)}(z)= \frac{\partial^n \psi(z)}{d z^n}, \quad H_n^{(r)}= \sum_{i=1}^n \frac{1}{i^r}, \quad \text{and}\quad H_n=H_n^{(1)}\non
\end{align}

The \textbf{one derivative term} either with $\partial_{\Delta_1},\partial_{\Delta_2}$ acting on the Beta function or $\partial_{\Delta}$ acting on the operator can also be evaluated. This term has explicit $\mathrm{Log}[-z_{12}\bar{z}_{12}\frac{\Lambda^2}{\mu^2}]$ which distinguishes it from other terms. This OPE can be evaluated using the manipulation with total derivative and contour integration. The total contribution with one derivative term is
\begin{align}
 \label{firstonederivative}
 &R^{a,+}_{k,-1}(z_1,\bar{z}_1)R^{b,+}_{l,-1}(z_2,\bar{z}_2)\sim \oint_{(k} \oint_{l)} O^{a,+}_{\Delta_1}(z_1,\bar{z}_1)  O^{b,+}_{\Delta_2}(z_2,\bar{z}_2)\supset  \nonumber\\
    & i \frac{(3g^2 N_c)}{8\pi^2}\frac{f^{abc}}{z_{12}}  \mathrm{Log}[-z_{12}\bar{z}_{12}\frac{\Lambda^2}{\mu^2}]
   \Bigg(H_{1-l} \left[R_{k+l-1,-1}^{c,+}+2 R_{k+l-1,-2}^{c,+} H_{-k-l+2}\right]\non\\
  &+H_{-k+1} \left[R_{k+l-1,-1}^{c,+}+2
   R_{k+l-1,-2}^{c,+} H_{-k-l+2}\right]-2 R_{k+l-1,-1}^{c,+} H_{-k-l+2}\non\\
  & -2 R_{k+l-1,-2}^{c,+} \left(H_{-k-l+2}\right){}^2-R_{k+l-1,-2}^{c,+}
   \left(H_{-k+1}\right){}^2-R_{k+l-1,-2}^{c,+} \left(H_{1-l}\right){}^2\non\\
   &-2 R_{k+l-1,-2}^{c,+} \psi ^{(1)}(-k-l+3)+R_{k+l-1,-2}^{c,+} \psi
   ^{(1)}(-k+2)\non\\
   &+R_{k+l-1,-2}^{c,+} \psi ^{(1)}(2-l)\Bigg)\frac{\Gamma (-k-l+3)} { \Gamma (2-l) \Gamma (-k+2)}   
 \end{align}

Hence, the total OPE of $R^{a,+}_{k,-1}(z_1,\bar{z}_1)R^{b,+}_{l,-1}(z_2,\bar{z}_2)$ is the \textbf{sum} of \eqref{firstzeroder}, \eqref{firsttwoder} and \eqref{firstonederivative}. In summary, this OPE receives contributions from both the tree level and the loop level. We will work out the special case of leading and subleading soft operators.
 \subsection*{Leading and subleading soft operators}
 For the leading soft theorem we have to set $k=l=1$, and then the OPE of leading soft operators will become
 \begin{align}
 \label{leading first1}
  & R^{a,+}_{1,-1}(z_1,\bar{z}_1)R^{b,+}_{1,-1}(z_2,\bar{z}_2) \sim  \oint_{(k=1} \oint_{l=1)}O^{a,+}_{\Delta_1}(z_1,\bar{z}_1)  O^{b,+}_{\Delta_2}(z_2,\bar{z}_2)\non\\  
   & \sim i \frac{f^{abc}}{z_{12}} R_{1,-1}^{c,+}(z_2,\bar{z}_2)+ i \frac{g^2 N_c}{8 \pi^2}\frac{f^{abc}}{z_{12}}  
\Bigg[-\frac{\pi^2}{12}\Bigg]  R_{1,-1}^{c,+}
-i \frac{g^2 N_c}{16 \pi^2}\frac{f^{abc}}{z_{12}}  
\Bigg[\mathrm{Log}^2[- z_{12}\bar{z}_{12} \frac{\Lambda^2}{\mu^2}]\Bigg]  R_{1,-1}^{c,+}\nonumber\\
&+ i \frac{(g^2 N_c)}{\pi^2}\frac{f^{abc}}{z_{12}}  \gamma^3 R_{1,-2}^{c,+}\big(1+\gamma\big)
 \end{align}
Here $\gamma$ is the Euler Gamma. In this example of the OPE of leading soft operators, we can observe its essential features. The OPE of simple poles in conformal soft operators results in conformal soft operators with higher poles. In the OPE given by \eqref{leading first1}, the origin of the first term is at the tree level, while the second, third, and fourth terms arise from loop corrections. The theory dependence can also be seen from the explicit factor of $N_c$ which represents the gauge group as $SU(N_c)$. \\
In the next example, we will consider the OPE of leading with subleading operators which amounts to $k=1,l=0$. 
\begin{align}
 \label{leading first12}
  & R^{a,+}_{1,-1}(z_1,\bar{z}_1)R^{b,+}_{0,-1}(z_2,\bar{z}_2) \sim  \oint_{(k=1} \oint_{l=0)}O^{a,+}_{\Delta_1}(z_1,\bar{z}_1)  O^{b,+}_{\Delta_2}(z_2,\bar{z}_2)\non\\  
   & \sim i \frac{f^{abc}}{z_{12}} R_{0,-1}^{c,+}(z_2,\bar{z}_2)+ i \frac{g^2 N_c}{8 \pi^2}\frac{f^{abc}}{z_{12}}  
\Bigg[\frac{1}{3}(1+\frac{n_s}{N_c}-\frac{n_f}{N_c})-\frac{\pi^2}{12}\Bigg]  R_{0,-1}^{c,+}
-i \frac{g^2 N_c}{16 \pi^2}\frac{f^{abc}}{z_{12}}  
\Bigg[\mathrm{Log}^2[- z_{12}\bar{z}_{12} \frac{\Lambda^2}{\mu^2}]\Bigg]  R_{0,-1}^{c,+}\nonumber\\
&+i \frac{g^2 N_c}{16 \pi^2}\frac{f^{abc} }{z_{12}^2}  \frac{1}{3}(1+\frac{n_s}{N_c}-\frac{n_f}{N_c})  \bar{z}_{12}  R_{0,-1}^{c,-}
+ i \frac{(g^2 N_c)}{\pi^2}\frac{f^{abc}}{z_{12}}  \gamma  (\gamma  (3+(\gamma -3) \gamma )-4) R_{0,-2}^{c,+}\nonumber\\
   &-i \frac{(3g^2 N_c)}{8\pi^2}\frac{f^{abc}}{z_{12}}  \mathrm{Log}[-z_{12}\bar{z}_{12}\frac{\Lambda^2}{\mu^2}]R_{0,-1}^{c,+}
 \end{align}
Here $\gamma$ is the Euler Gamma. This OPE has loop corrections, theory dependence, and the presence of higher poles from the OPE of simple pole operators. 

 \subsection{Second type OPE}
 Now we are going to evaluate the second type of OPE $R^{a,+}_{k,-2}(z_1,\bar{z}_1)R^{b,+}_{l,-2}(z_2,\bar{z}_2)$ which comes strictly from the loop corrections. First, the term without any derivative in \eqref{ope gluons positive both raw} \textbf{won't contribute}. Next, the term with two derivatives in total can be written as


 \begin{eqnarray}
 \label{twodertwo}
 && R^{a,+}_{k,-2}(z_1,\bar{z}_1)R^{b,+}_{l,-2}(z_2,\bar{z}_2)\sim\oint_{(k} \oint_{l)} (\Delta_1-k)(\Delta_2-l)O^{a,+}_{\Delta_1}(z_1,\bar{z}_1)  O^{b,+}_{\Delta_2}(z_2,\bar{z}_2)\supset  \nonumber\\
    &&  i \frac{(g^2 N_c)}{8\pi^2}\frac{f^{abc}}{z_{12}}   
   \left(9 R_{k+l-1,-2}^{c,+} \left[-2 H_{-k-l+2}+H_{-k+1}+H_{1-l}\right]-8 R_{k+l-1,-1}^{c,+}\right)\frac{ \Gamma (-k-l+3)}{\Gamma
   (2-l) \Gamma (-k+2)}\non\\
 \end{eqnarray}
We observe that the OPE of the conformal operator with double poles yields a conformal operator with both simple and double poles. Next, the term with single derivative which contains an explicit factor of $\mathrm{log}$ will give the following OPE
 \begin{eqnarray}
 \label{twoderone}
 && R^{a,+}_{k,-2}(z_1,\bar{z}_1)R^{b,+}_{l,-2}(z_2,\bar{z}_2)\sim\oint_{(k} \oint_{l)} (\Delta_1-k)(\Delta_2-l)O^{a,+}_{\Delta_1}(z_1,\bar{z}_1)  O^{b,+}_{\Delta_2}(z_2,\bar{z}_2)  \nonumber\\
  &&\supset i  \frac{(g^2 N_c)}{2\pi^2} \frac{f^{abc}}{z_{12}} \mathrm{Log}[-z_{12}\bar{z}_{12}\frac{\Lambda^2}{\mu^2}] \Bigg(\frac{ \Gamma (-k-l+3)}{\Gamma (2-l) \Gamma (-k+2)}\Bigg) R_{k+l-1,-2}^{c,+}
 \end{eqnarray}
 Interestingly, this OPE only produces conformal operators having double poles. Therefore, the complete OPE of $R^{a,+}_{k,-2}(z_1,\bar{z}_1)R^{b,+}_{l,-2}(z_2,\bar{z}_2)$ is the sum of \eqref{twodertwo} and \eqref{twoderone}.
 \subsection*{Leading and subleading soft operators}
 For the leading soft operators, we take $k=1,l=1$
 \begin{eqnarray}
 \label{twodertwo}
 && R^{a,+}_{1,-2}(z_1,\bar{z}_1)R^{b,+}_{1,-2}(z_2,\bar{z}_2)\sim\oint_{(k=1} \oint_{l=1)} (\Delta_1-k)(\Delta_2-l)O^{a,+}_{\Delta_1}(z_1,\bar{z}_1)  O^{b,+}_{\Delta_2}(z_2,\bar{z}_2)  \nonumber\\
    && \sim -i \frac{(g^2 N_c)}{\pi^2}\frac{f^{abc}}{z_{12}}   
   R_{1,-1}^{c,+}+i  \frac{(g^2 N_c)}{2\pi^2} \frac{f^{abc}}{z_{12}} \mathrm{Log}[-z_{12}\bar{z}_{12}\frac{\Lambda^2}{\mu^2}]R_{1,-2}^{c,+}\non\\
 \end{eqnarray}
 Next, we are going to calculate the OPE of leading with subleading operators ($k=1,l=0$).
 \begin{eqnarray}
 \label{twodertwo}
 && R^{a,+}_{1,-2}(z_1,\bar{z}_1)R^{b,+}_{0,-2}(z_2,\bar{z}_2)\sim\oint_{(k=1} \oint_{l=0)} (\Delta_1-k)(\Delta_2-l)O^{a,+}_{\Delta_1}(z_1,\bar{z}_1)  O^{b,+}_{\Delta_2}(z_2,\bar{z}_2)  \nonumber\\
    && \sim i \frac{(g^2 N_c)}{8 \pi^2}\frac{f^{abc}}{z_{12}} \Bigg(-9  R_{1,-2}^{c,+}-8   
   R_{1,-1}^{c,+}\Bigg)+i  \frac{(g^2 N_c)}{2\pi^2} \frac{f^{abc}}{z_{12}} \mathrm{Log}[-z_{12}\bar{z}_{12}\frac{\Lambda^2}{\mu^2}]R_{1,-2}^{c,+}\non\\
 \end{eqnarray}
 In these examples, we observe the effects of loop corrections and the presence of the gauge group factor $N_c$. The OPE of conformal operators with double poles leads to the appearance of conformal operators with both simple and double poles.
 \subsection{Third type OPE}
Next, we are going to evaluate the mixed OPE between the conformal soft operators having the double and simple poles $R^{a,+}_{k,-2}(z_1,\bar{z}_1)R^{b,+}_{l,-1}(z_2,\bar{z}_2)$. This OPE has contributions from both trees and loops. We will separate the contributions of trees and loops as the expressions are cumbersome to write in a single equation.
The tree-level contributions can be written as before
\begin{align}
\label{thirdzeroder}
    & R^{a,+}_{k,-2}(z_1,\bar{z}_1)R^{b,+}_{l,-1}(z_2,\bar{z}_2) \sim  \oint_{(k} \oint_{l)}  (\Delta_1-k)    O^{a,+}_{\Delta_1}(z_1,\bar{z}_1)  O^{b,+}_{\Delta_2}(z_2,\bar{z}_2)\non\\
    & \sim i \frac{f^{abc}}{z_{12}}\frac{(2-k-l)!}{(1-k)!(1-l)!} R_{k+l-1,-1}^{c,+}(z_2,\bar{z}_2)\non\\
 &   + i \frac{g^2 N_c}{8 \pi^2}\frac{f^{abc}}{z_{12}}  
\Bigg[\frac{1}{3}(1+\frac{n_s}{N_c}-\frac{n_f}{N_c})\frac{(-k-l)!}{(-k)!(-l)!}
-\frac{\pi^2}{12}\frac{(2-k-l)!}{(1-k)!(1-l)!}\Bigg] 
 R_{k+l-1,-1}^{c,+}\non\\
&-i \frac{g^2 N_c}{16 \pi^2}\frac{(2-k-l)!}{(1-k)!(1-l)!}\frac{f^{abc}}{z_{12}}  
\Bigg[\mathrm{Log}^2[- z_{12}\bar{z}_{12} \frac{\Lambda^2}{\mu^2}]\Bigg] 
 R_{k+l-1,-1}^{c,+}\non\\
&+i \frac{g^2 N_c}{16 \pi^2}\frac{f^{abc} }{z_{12}^2}  \frac{1}{3}(1+\frac{n_s}{N_c}-\frac{n_f}{N_c})  
  \frac{(-k-l)!}{(-k)!(-l)! }\bar{z}_{12}  R_{k+l-1,-1}^{c,-}\non\\  
  \end{align}
The contribution of two derivative terms to the OPE can be written as
 \begin{align}
 \label{thirdtwoder}
 &R^{a,+}_{k,-2}(z_1,\bar{z}_1)R^{b,+}_{l,-1}(z_2,\bar{z}_2)\sim\oint_{(k} \oint_{l)} (\Delta_1-k)O^{a,+}_{\Delta_1}(z_1,\bar{z}_1)  O^{b,+}_{\Delta_2}(z_2,\bar{z}_2)\supset  i \frac{(g^2 N_c)}{8\pi^2}\frac{f^{abc}}{z_{12}}   
   \nonumber\\
&  \Bigg(5 \psi ^{(0)}(2-l) \Big[R_{k+l-1,-1}^{c,+}+2 R_{k+l-1,-2}^{c,+} \psi ^{(0)}(-k-l+3)\Big]-\psi ^{(0)}(-k+2)\non\\
 & \Big[9
   R_{k+l-1,-1}^{c,+}+32 R_{k+l-1,-2}^{c,+} \psi ^{(0)}(-k-l+3)\Big]+4 R_{k+l-1,-1}^{c,+} \psi ^{(0)}(-k-l+3)\non\\
   &+11 R_{k+l-1,-2}^{c,+} \psi
   ^{(0)}(-k-l+3)^2+11 R_{k+l-1,-2}^{c,+} \psi ^{(1)}(-k-l+3)\non\\
   &+16 R_{k+l-1,-2}^{c,+} \psi ^{(0)}(-k+2)^2-16 R_{k+l-1,-2}^{c,+} \psi ^{(1)}(-k+2)-5
   R_{k+l-1,-2}^{c,+} \psi ^{(0)}(2-l)^2\non\\
   &+5 R_{k+l-1,-2}^{c,+} \psi ^{(1)}(2-l)\Bigg) \Bigg( -\frac{ \Gamma (-k-l+3)  }{\Gamma (2-l) \Gamma (-k+2)}\Bigg)  \non\\
 \end{align}

The contribution of one derivative term to the OPE can be written as
\begin{align}
\label{thirdoneder}
& \oint_{(k} \oint_{l)} (\Delta_1-k)O^{a,+}_{\Delta_1}(z_1,\bar{z}_1)  O^{b,+}_{\Delta_2}(z_2,\bar{z}_2)\supset  i \frac{(g^2 N_c)}{8\pi^2}\frac{f^{abc}}{z_{12}}   
   \mathrm{Log}[-z_{12}\bar{z}_{12}\frac{\Lambda^2}{\mu^2}]  \non\\
   & \frac{ \Gamma (-k-l+3) }{\Gamma
   (2-l) \Gamma (-k+2)}\Bigg(R_{k+l-1,-1}^{c,+}+R_{k+l-1,-2}^{c,+} \Big[4 H_{-k-l+2}-5 H_{-k+1}+H_{1-l}\Big]\Bigg)\non\\
\end{align}

Therefore, the OPE of $R^{k,a}_{-2}(z_1,\bar{z}_1)R^{l,b}_{-1}(z_2,\bar{z}_2)$ is the sum of \eqref{thirdzeroder}, \eqref{thirdtwoder} and \eqref{thirdoneder}. \\
 \subsection*{Leading and subleading soft operators}
 First we will start with OPE of leading soft operators having $k=l=1$.
 \begin{align}
\label{thirdzeroder leading}
    & R^{a,+}_{1,-2}(z_1,\bar{z}_1)R^{b,+}_{1,-1}(z_2,\bar{z}_2) \sim  \oint_{(k=1} \oint_{l=1)}  (\Delta_1-k)    O^{a,+}_{\Delta_1}(z_1,\bar{z}_1)  O^{b,+}_{\Delta_2}(z_2,\bar{z}_2)\non\\
    &  \sim i \frac{f^{abc}}{z_{12}} R_{1,-1}^{c,+}(z_2,\bar{z}_2)+ i \frac{g^2 N_c}{8 \pi^2}\frac{f^{abc}}{z_{12}}  
\Bigg[-\frac{\pi^2}{12}\Bigg]  R_{1,-1}^{c,+}
-i \frac{g^2 N_c}{16 \pi^2}\frac{f^{abc}}{z_{12}}  
\Bigg[\mathrm{Log}^2[- z_{12}\bar{z}_{12} \frac{\Lambda^2}{\mu^2}]\Bigg]  R_{1,-1}^{c,+}\nonumber\\
&  + i \frac{(g^2 N_c)}{8\pi^2}\frac{f^{abc}}{z_{12}}   
   \mathrm{Log}[-z_{12}\bar{z}_{12}\frac{\Lambda^2}{\mu^2}] R_{1,-1}^{c,+}
    \end{align}
    The OPE of leading with subleading operators ($k=1,l=0$) can be written as
 \begin{align}
\label{thirdzeroder leading sub}
    & R^{a,+}_{1,-2}(z_1,\bar{z}_1)R^{b,+}_{0,-1}(z_2,\bar{z}_2) \sim  \oint_{(k=1} \oint_{l=0)}  (\Delta_1-k)    O^{a,+}_{\Delta_1}(z_1,\bar{z}_1)  O^{b,+}_{\Delta_2}(z_2,\bar{z}_2)\non\\
   & \sim i \frac{f^{abc}}{z_{12}} R_{0,-1}^{c,+}(z_2,\bar{z}_2)+ i \frac{g^2 N_c}{8 \pi^2}\frac{f^{abc}}{z_{12}}  
\Bigg[\frac{1}{3}(1+\frac{n_s}{N_c}-\frac{n_f}{N_c})-\frac{\pi^2}{12}\Bigg]  R_{0,-1}^{c,+}
-i \frac{g^2 N_c}{16 \pi^2}\frac{f^{abc}}{z_{12}}  
\Bigg[\mathrm{Log}^2[- z_{12}\bar{z}_{12} \frac{\Lambda^2}{\mu^2}]\Bigg]  R_{0,-1}^{c,+}\nonumber\\
&+i \frac{g^2 N_c}{16 \pi^2}\frac{f^{abc} }{z_{12}^2}  \frac{1}{3}(1+\frac{n_s}{N_c}-\frac{n_f}{N_c})  \bar{z}_{12}  R_{0,-1}^{c,-}
- 9i \frac{(g^2 N_c)}{8\pi^2}\frac{f^{abc}}{z_{12}}  
    R_{0,-1}^{c,+}\nonumber\\
   &+i \frac{(g^2 N_c)}{8\pi^2}\frac{f^{abc}}{z_{12}}  \mathrm{Log}[-z_{12}\bar{z}_{12}\frac{\Lambda^2}{\mu^2}](R_{0,-1}^{c,+}+5 R_{0,-2}^{c,+})
    \end{align}   
 \textbf{Summary:-} In this section, we found that the OPE of conformally soft gluons has simple and double poles. We worked out an explicit example of leading and subleading soft gluon operators. The essential features of these OPEs are the \textbf{mixing} of simple pole and double pole operators. This mixing comes essentially from loop corrections. The theory dependence can also be observed from the presence of factors like $N_c,n_s,n_f$.

\section{Loop level graviton OPE}
\label{section6}
In the gravitational scattering, the leading collinear limit remains all loop exact \cite{Bern:1998sv}. Hence the graviton-graviton OPE won't get modified. The proof of cancellation of collinear divergence in gravity is established in \cite{Akhoury:2011kq}. The intuitive idea is the graviton interaction vertices come with two powers of momenta, which "softens" the divergence. This fact is true irrespective of the signature of spacetime. The OPE of two gravitons remains unchanged and it is given by \cite{Pate:2019lpp,Guevara:2021abz}\footnote{To avoid any subtlety, we are making a statement only for the OPE of positive helicity gravitons. The OPE for opposite helicity gravitons is left for future exploration.}
\begin{eqnarray}
   G^+_{\Delta_1}(z_1,\bar{z}_1)G^+_{\Delta_2}(z_2,\bar{z}_2)\sim -\frac{\kappa}{2 z_{12}}\sum_{n=0}^{1-k} B(\Delta_1-1+n,\Delta_2-1)\frac{(\bar{z}_{12})^{n+1}}{n!} \bar{\partial}^n G^+_{\Delta_1+\Delta_2}(z_2,\bar{z}_2) \non\\
 \end{eqnarray}
The effects of the loop will make simple poles in the soft operators to higher poles. At one loop level, We will have simple poles and 2nd order poles in the expansion of the soft operator. The higher poles come with an additional $\kappa^2=32 \pi G$ relative to the simple (tree) poles. The definition of conformally soft graviton with their modes can be written as \cite{Guevara:2021abz}
\begin{eqnarray}
 &&H_{k,-1}(z,\bar{z})    =\oint_k \frac{d \Delta}{2 \pi i} G_{\Delta}(z,\bar{z}) , \quad  H_{k,-2}(z,\bar{z})    =\oint_k \frac{d \Delta}{2 \pi i}(\Delta-k) G_{\Delta}(z,\bar{z})\nonumber\\
   &&H_{k,-1}(z,\bar{z})= \sum_{n=\frac{k-2}{2}}^{\frac{2-k}{2}} \frac{H^k_{n,-1}(z)}{\bar{z}^{n+\frac{k-2}{2}}},\quad  H_{k,-2}(z,\bar{z})= \sum_{n=\frac{k-2}{2}}^{\frac{2-k}{2}} \frac{H^k_{n,-2}(z)}{\bar{z}^{n+\frac{k-2}{2}}}
\end{eqnarray}
For the second case $ H_{k,-2}(z,\bar{z})$ the conformal dimension $k\in \{0,-1,-2,\cdots\}$. The reason lies in the fact that the leading soft graviton theorem doesn't get loop corrected.
We will start with the OPE of simple pole operators following \cite{Guevara:2021abz} 
\begin{eqnarray}
   H^+_{k,-1}(z_1,\bar{z}_1)H^+_{l,-1}(z_2,\bar{z}_2)\sim -\frac{\kappa}{2 z_{12}}\sum_{n=0}^{1-k} \frac{(2-k-l-n)!}{(1-k-n)!(1-l)!}\frac{(\bar{z}_{12})^{n+1}}{n!} \bar{\partial}^n H^+_{k+l,-1}(z_2,\bar{z}_2) \non\\
\end{eqnarray}
For higher poles conformal operators we have an additional OPE
\begin{eqnarray}
   H^+_{k,-2}(z_1,\bar{z}_1)H^+_{l,-1}(z_2,\bar{z}_2)\sim -\frac{\kappa}{4 z_{12}}\sum_{n=0}^{1-k} \frac{(2-k-l-n)!}{(1-k-n)!(1-l)!}\frac{(\bar{z}_{12})^{n+1}}{n!} \bar{\partial}^n H^+_{k+l,-2}(z_2,\bar{z}_2) \non\\
\end{eqnarray}
The OPE of $ H^+_{k,-2}(z_1,\bar{z}_1)H^+_{l,-2}(z_2,\bar{z}_2)$ vanishes.\\

Notice that we have an additional factor of $\frac{1}{2}$ as compared to tree-level OPE. The origin lies in the symmetrization of the contour. Again we can write down the commutators of the modes similar to the \cite{Guevara:2021abz}
\begin{eqnarray}
\label{com1}
\Big[H^k_{m,-1},H^l_{n,-1}\Big]=-\frac{\kappa}{2} [n(2-k)-m(2-l)]\frac{( \frac{2-k}{2}-m+\frac{2-l}{2}-n-1)!}{( \frac{2-k}{2}-m)!(\frac{2-l}{2}-n)!}
   \frac{(\frac{2-k}{2}+m +\frac{2-l}{2}+n-1)!}{(\frac{2-k}{2}+m)!( \frac{2-l}{2}+n)!} H^{k+l}_{m+n,-1}\non\\
\end{eqnarray}
The other commutator can be written as
\begin{eqnarray}
\label{com2}
\Big[H^k_{m,-2},H^l_{n,-1}\Big]=-\frac{\kappa}{4} [n(2-k)-m(2-l)]\frac{( \frac{2-k}{2}-m+\frac{2-l}{2}-n-1)!}{( \frac{2-k}{2}-m)!(\frac{2-l}{2}-n)!}
   \frac{(\frac{2-k}{2}+m +\frac{2-l}{2}+n-1)!}{(\frac{2-k}{2}+m)!( \frac{2-l}{2}+n)!} H^{k+l}_{m+n,-2}\non\\
\end{eqnarray}
and after doing the light transform as in \cite{Strominger:2021lvk}\footnote{Strominger \cite{Strominger:2021lvk} has added the central generator with $k=2$ or $p=1$ with the corresponding mode as $H^2_0(z)$ which is a current with $(h,\bar{h})=(2,0)$. It is central in supertranslation algebra \eqref{com1} with $k=1,l=1$. In our case, we have the mode algebra as \eqref{com2}. The commutator with $k=0,l=1$ gives a graviton mode $H^1_{m+n,-2}$. Hence we can also central generators $H^k_{m,-2}$ with $k=1,m= \pm \frac{1}{2}$ (not to confused with leading soft graviton operator which has $k=1$ but doesn't receive any loop corrections). We are tempted to add another central generator $H^2_{0,-2}$ but it is not being inferred by the OPE \eqref{com2}. Hence, we won't add it to the OPE.} (notice the dependence on $\kappa$)
\begin{eqnarray}
    w_{n,-1}^p=\frac{1}{\kappa} (p-n-1)!(p+n-1)! H_{n,-1}^{-2p+4}, \quad p\in \{1,\frac{3}{2},2,\frac{5}{2} \cdots\}\\
     w_{n,-2}^p=\frac{1}{\kappa} (p-n-1)!(p+n-1)! H_{n,-2}^{-2p+4}, \quad p\in \{\frac{3}{2},2,\frac{5}{2},  \cdots\}
\end{eqnarray}
Then the above commutator \eqref{com1} can be written as
\begin{eqnarray}
\label{com1w}
    [w_{m,-1}^p,w_{n,-1}^q]=[m(q-1)-n(p-1)]w_{m+n,-1}^{p+q-2}
\end{eqnarray}
The other commutator \eqref{com2} can be written  as
\begin{eqnarray}
\label{com2w}
    [w_{m,-2}^p,w_{n,-1}^q]=\frac{1}{2}[m(q-1)-n(p-1)]w_{m+n,-2}^{p+q-2}
\end{eqnarray}
The central generator $w_{m,-2}^p$ with $p=\frac{3}{2}$ commutes with all other generators. 
 The modes labelled by $m$ are restricted as $\frac{k-2}{2}\leq m \leq \frac{2-k}{2}$. Hence the algebra presented in \eqref{com1w} is the wedge sub-algebra of $w_{1+\infty}$ algebra. The algebra in \eqref{com2w} is also a wedge subalgebra of $w_{\infty}$ algebra.

This concludes our discussion on the graviton OPE at the one loop level. The higher loop let's say 2-loop contribution won't change the subleading soft graviton theorem but it can change the sub-subleading soft theorem. Up to 2-loop the sub-subleading soft theorem has a dependence like $\omega \mathrm{Log}^2[\omega]$. This will translate to the 3rd order pole but again at $\Delta=-1$. Then again we have a new conformal soft graviton operator $H_{k,-3}$ but $k \in \{ -1,-2, -3, \cdots \}$. 
The new commutator will be \footnote{We can add central generators as $w_{m,-3}^2$, which is needed from the commutator of $w_{m,-3}^{\frac{5}{2}}$ and $w_{m,-1}^{\frac{3}{2}}$.  }
\begin{eqnarray}
\label{com3w}
    [w_{m,-3}^p,w_{n,-1}^q]=\frac{1}{2}[m(q-1)-n(p-1)]w_{m+n,-3}^{p+q-2}, \quad (p+q-2) \in \{\frac{5}{2},3, \frac{7}{2}, \cdots\}. 
\end{eqnarray}
So, up to 2 loop order, the resulting symmetry algebra will become the sum of the above algebra as
\begin{eqnarray}
\label{algebrawfinal}
[w_{m,-1}^{p_1}+w_{m,-2}^{p_2}+w_{m,-3}^{p_3},w_{n,-1}^q]=&&\Bigg([m(q-1)-n(p_1-1)]w_{m+n,-1}^{p_1+q-2}+\frac{1}{2}[m(q-1)-n(p_2-1)]w_{m+n,-2}^{p_2+q-2}\non\\
&&+\frac{1}{2}[m(q-1)-n(p_3-1)]w_{m+n,-3}^{p_3+q-2}\Bigg),
\end{eqnarray}
$p_i+q-2 $ should belong to the appropriate sector as calculated above.  This structure will get generalized to higher loops.\\

\textbf{Conclusions:-} In this section, we have seen that the wedge subalgebra of $w_{1+\infty}$ algebra found at tree level get duplicated. We found an anthoer set of wedge subalgebra of $w_{\infty}$ algebra at  one loop level. A similar structure persists at a higher loop as well. But we don't think it is $W_{1+\infty}$- the quantum deformation of $w_{1+\infty}$.

\section{Discussions and Outlook}
In this article, we have initiated the study of loop-corrected celestial soft gluon and graviton algebra. We started with the generic one loop gluon OPE found by \cite{Bhardwaj:2022anh} and made the operator soft by taking conformal dimension $\Delta\in\{1,0,-1,\cdots\}$. One can ask the following question, what if we study the double soft theorem and make them collinear? These two limits commute or not. We believe these two limits commute. Here we will present an argument in support of this belief. The argument comes from an existence of CCFT at celestial sphere for soft gluons.

The conformal operators $O_{\Delta}(z,\bar{z})$ live on the celestial sphere. The soft operator having dimension $\Delta\in\{1,0,-1,\cdots\}$ is one of them. The collinear limit translates on to the the OPE limit. And OPE of two operators do commute \footnote{Here we are assuming the conformal operators are bosonic.}.
So, we need a CCFT description for soft gluons. Happily, the CCFT description for QCD in the soft and collinear sectors is first presented by Magnea \cite{Magnea:2021fvy}. This CCFT is just the Lie algebra valued free fields (see section-4 of \cite{Magnea:2021fvy}). To really test the CFT description, Magnea calculated the splitting anomalous dimension from free field OPE. This calculation mimics the bulk gauge theory calculation of splitting functions and its anomalous dimensions. The answer one gets from the CFT description matches with the bulk gauge theory calculations. It also establishes the fact that the collinear limit depends only on the particles involved in the splitting and color correlations with other particles vanish in the leading order. The free field OPE in CFT encodes the collinear factorization theorem \footnote{The factorization theorem states that amplitudes in QCD factorize as hard-collinear-soft. Hence, the study of the soft theorem can be done independently of the collinear limit and vice versa (See \cite{Herzog:2023sgb} for some recent explicit results).} of gauge theory (see eq. 229 of \cite{Feige:2014wja}). One can do a similar analysis in gravity following \cite{White:2011yy}. We will leave the detailed analysis for future work.\\

In summary, the soft gluon operator gets the loop corrections. The tree level operator when changed to Mellin basis only has simple poles in the $\Delta$ plane. But the Loop effects contribute as $g^3_{YM}\frac{\mathrm{Log}[\omega]}{\omega}$. The loop correction has an extra $g^2_{YM}$ compared to the tree-level operator. The $\frac{\mathrm{Log}[\omega]}{\omega}$, when expressed in the Mellin basis, exhibits double poles, precisely at the same locations as those observed at the tree level ($\Delta=k$). We can see even though loop correction changed the nature of poles but still, soft operators still lie on the same discrete points $\Delta=\{1,0,-1,\cdots\}$. It lies exactly at the same place even if we keep on adding the higher loop corrections.\\

With this input, we obtained the OPE of conformal soft operators. We have two types of conformal soft operators $R^a_{k,-1}$ and $R^a_{k,-2}$ depending on simple poles or higher poles. We obtained their OPE in section \ref{section4}. One needs to be very careful about the addition of $SL(2,R)_R$ descendants as loop level collinear limit is more subtle in split signature. Luckily for gravitons, the leading order-splitting functions obtained at the tree level are all loop exact. But subleading soft gravitons have loop corrections (one loop exact), which translates to the second-order poles in the $\Delta$ plane. We found the tree and loop-corrected conformal soft operators and their OPE. The loop correction added a new set of  wedge subalgebra of $w_{\infty}$ algebra to the existing wedge subalgebra of $w_{1+\infty}$. The higher loop correction will follow suit. The algebra written in \eqref{algebrawfinal} looks like some extension of $w_{1+\infty}$ algebra than the qunatum $W_{1+\infty}$ algebra.\\

There are many future directions we would like to follow.
\begin{itemize}
\item The collinear splitting functions for gluons and gravitons are known at all loop order in planar $\mathcal{N}=4$ SYM (see \cite{Bhardwaj:2022anh}). One can do a similar analysis as done in this paper to find the mode algebras of the conformal soft operators.
\item The loop corrected stress tensor and their OPE with other operators has been obtained in \cite{Pasterski:2022djr,Donnay:2021wrk,Donnay:2022hkf}). Recently, Agrawal et al.\cite{Agrawal:2023zea} have found the connection between the logarithmic soft theorems and superrotation Ward identity. One can speculate whether higher soft theorems can be recast in terms of Ward identities or not. It would be very interesting to understand the BMS flux algebra for higher logarithmic soft theorems. This will complement the analysis of graviton OPE done in this paper. 
\item It has been found in \cite{Freidel:2022skz} that the operator lying on discrete conformal dimensions more precisely the memory and goldstone modes form the basis of conformal primary wavefunctions provided the news and shear follow certain fall-off conditions (gravitational signal belonging to Schwartz space). This sounds a little problematic from the view of gravitation tail memory which is not exponentially suppressed at late time and hence doesn't belong to the Schwartz space \cite{Ghosh:2021bam}. This can be traced back to the relationship between gravitational waveform and soft theorem. As the soft theorem gets loop correction, we can see the gravitational memory also gets corrected. We wish to understand the completeness of the memory and goldstone modes at the loop level.
\item There is a well-developed theory effective theory for soft and collinear modes of QCD and gravity called soft-collinear effective theory (SCET). We think SCET can be very helpful in understanding the loop correction of celestial algebra. In the factorization theorem of \cite{Feige:2014wja}, the soft theorem is obtained using the smeared Wilson lines.
\begin{eqnarray}
    Y^{\dagger}_j(x)= P \, \, \mathrm{exp}\Bigg[i g \int _0^{\infty} ds\,  n_j. A(x^{\nu}+s n_j^{\nu})e^{-\epsilon s}\Bigg] 
\end{eqnarray}
when $n_j$ is along the null direction, then the Wilson line looks more like light-ray operators. The soft theorem can be obtained after expanding the matrix elements of these Wilson lines. This will strengthen the relationship between light ray operators and soft theorems (see a recent discussion about light ray operators and soft theorem \cite{Hu:2022txx,Hu:2023geb}).
\end{itemize}

\label{outlook}
\begin{acknowledgments}

We are indebted to J.~Bhambure, R.~Bhardwaj, Y.~Hu, B.~Sahoo, G. Stermann, Y.P.~Wang, for helpful discussions. We would like to especially thank R.~Bhardwaj for going through the previous draft and pointing out errors. We would also like to thank G. Stermann for important comments on the draft.  The work of HK is supported by NSF grant PHY 2210533.

\end{acknowledgments}
\appendix
\section{Some properties of Mellin transform}
\label{Mellin requirement}
In this section, we will give some reviews of the Mellin transform and its inverse properties. More details and proofs of the theorem can be found in \cite{Guevara:2019ypd}. 

The Mellin transform is defined by
\begin{eqnarray}
\label{mellin}
    \tilde{A}(\Delta)=\int_0^{\infty} \frac{d \omega}{\omega} \omega^{\Delta} A(\omega)
\end{eqnarray}
If  $A(\omega)$ is integrable then the only singularities of the above integral are at $\omega=0\, \mathrm{and}\,\,\infty$. Hence $A(\omega)$ will admit an asymptotic expansion 
\begin{eqnarray}
\label{log expansion}
    A(\omega) \xrightarrow {\omega \rightarrow 0 }  \sum_{p>i\geq -a} c_{i,j} \omega^i \mathrm{Log}^j[\omega]+O(\omega^p)\non\\
  A(\omega)\xrightarrow {\omega \rightarrow \infty }\sum_{q>i\geq -b} d_{i,j} \omega^i \mathrm{Log}^j[\omega]+O(\omega^q)  \end{eqnarray}
  The convergence of \eqref{mellin} requires $a<b$. This defines a fundamental strip $(a,b)$ for $\tilde{A}(\Delta)$. The $c_{i,j}$ are some constants but they could be a function of coupling constants for the soft theorems.\\
  For the study of the soft theorem, we always have $a=1$. The behavior of amplitude at $\omega \rightarrow \infty$ can be constrained using Froissart bound. But this bound is valid only in certain kinematic regimes. Nevertheless, the fundamental strip $(a,b)$ is not empty. For this strip $\tilde{A}$ is analytic in the complex $\Delta$ plane. \\
  The inverse Mellin transform can also be defined as
  \begin{eqnarray}
      A(\omega)=\frac{1}{2 \pi i} \int_{c-i \infty}^{c+i \infty} \frac{d \Delta}{\omega^{\Delta}} \tilde{A}(\Delta)
  \end{eqnarray}
  here $\tilde{A}(\Delta)$ be integrable on the imaginary line $c+i \mathbb{R}$, and $a<c<b$.\\
  
 In soft theorem analysis, we have an asymptotic expansion near $\omega \rightarrow 0$. Using this asymptotic expansion near $\omega \rightarrow 0$, we will establish some of the properties of $\tilde{A}(\Delta)$. 
 
  \begin{theorem}
 Let $\tilde{A}(\Delta)$ be the Mellin transform of $A(\omega)$ with a non-empty fundamental strip as $(a, b)$. And if $A(\omega)$ admits an asymptotic expansion in the form of \eqref{log expansion} around $\omega\rightarrow 0$ as well as at $\omega\rightarrow \infty$, then one can continue $\tilde{A}(\Delta)$  to a meromorphic function in the strip $(-p,-q)$, with this singular expansion
\begin{eqnarray}
\label{log mellin}
    \tilde{A}(\Delta)\asymp \sum_{i\geq -a,j\geq 0} c_{i,j} \frac{(-1)^j j!}{(\Delta+i)^{j+1}}+\sum_{i\leq -b,j\geq 0} d_{i,j} \frac{(-1)^j j!}{(\Delta+i)^{j+1}}
\end{eqnarray}
\end{theorem} 
The theorem also has a converse (called inverse mapping theorem), which states that the asymptotic form of $f(\omega)$ can be obtained from the pole structure of $\tilde{f}(\Delta)$ under suitable conditions. We are not going to need the details of the converse. Interested readers can look at \cite{Guevara:2019ypd}.\\

In our discussion of the soft theorem in section \ref{soft operator}, we have established that the amplitudes have an asymptotic expansion near $\omega \rightarrow 0$ as $\sum c_{i,j}\omega^i \mathrm{Log}^j[\omega]$. It is exactly like the expansion in \eqref{log expansion} with an appropriate choice of $p$ and $a$. Hence, we can use the   
theorem \eqref{log mellin}. On the $\Delta$ basis, the soft expansion will become
\begin{eqnarray}
\sum c_{i,j}\omega^i \mathrm{Log}^j[\omega] \rightarrow \sum c_{i,j}\frac{(-1)^jj!}{(\Delta+i)^{j+1}}    
\end{eqnarray}

One immediate consequence that can be reached here is the soft operators still line on the discrete series $\Delta=1- \mathbb{Z}_+$ but with higher order poles. The order of higher poles is given by the power of ($\mathrm{log}$) in the asymptotic expansion. It is very important to note that the whole $\Delta$ dependence of soft operators is written as $\frac{(-1)^jj!}{(\Delta+i)^{j+1}}$ and $c_{i,j}$ are just some coefficients which could be a function of coupling and other data of theory like gauge group, matter, etc. \\

Building from this intuition, we will write the soft operator as an expansion in the $\Delta$ plane. The soft operator can have celestial coordinate dependence ($z,\bar{z}$). 
\begin{eqnarray}
      O_{\Delta}(z,\bar{z})\sim c_{i,0}\frac{O^{-1}(z,\bar{z})}{(\Delta+i)}+c_{i,1}\frac{O^{-2}(z,\bar{z})}{(\Delta+i)^2}+\cdots
  \end{eqnarray}
  One can explicitly find the coupling constant dependence of these soft operators. The tree level soft operator has $g_{YM}$ dependence, while one loop level has $g_{YM}^3$ dependence. We have incorporated these coupling constants information in $c_{i,j}$. \\
  
  In the graviton case, the leading operator doesn't receive loop correction, hence we have $i=-1,j=0$. The subleading operator has both tree and loop parts. So, we have both $c_{i,0}$ and $c_{i,1}$ terms. In the gluon case, the leading soft theorem also gets corrected. In this article, we have just focused on one loop contribution, so the leading operator has both $c_{i,0}$ and $c_{i,1}$ terms. The higher loop will produce higher poles.
  \section{One loop soft gluon theorem}
  \label{soft gluon}
  In this section, we are going to sketch the one-loop soft gluon theorem. The leading soft gluon theorem has been studied very extensively (see \cite{Catani:2000pi} and the references therein). The soft factor is calculated in dimensional regularization and the leading term diverges as $\frac{1}{\epsilon^2}$. The Grammer-Yennie prescription (\cite{PhysRevD.8.4332}) allows us to separate the soft divergence as they tend to exponentiate. The divergences are captured by $K$ gluon propagator and we will simply factor out these divergences from $\mathcal{A}_{n+1}$ and $\mathcal{A}_n$. The remaining $G$ gluon piece will give us a finite contribution to the amplitude. Hence one should study the soft theorem for these finite amplitudes (see \cite{Sen:1981sd} and the reference therein for K-G decomposition for gluons).\\
  
\begin{center}
\begin{figure}
	\includegraphics[scale=0.60]{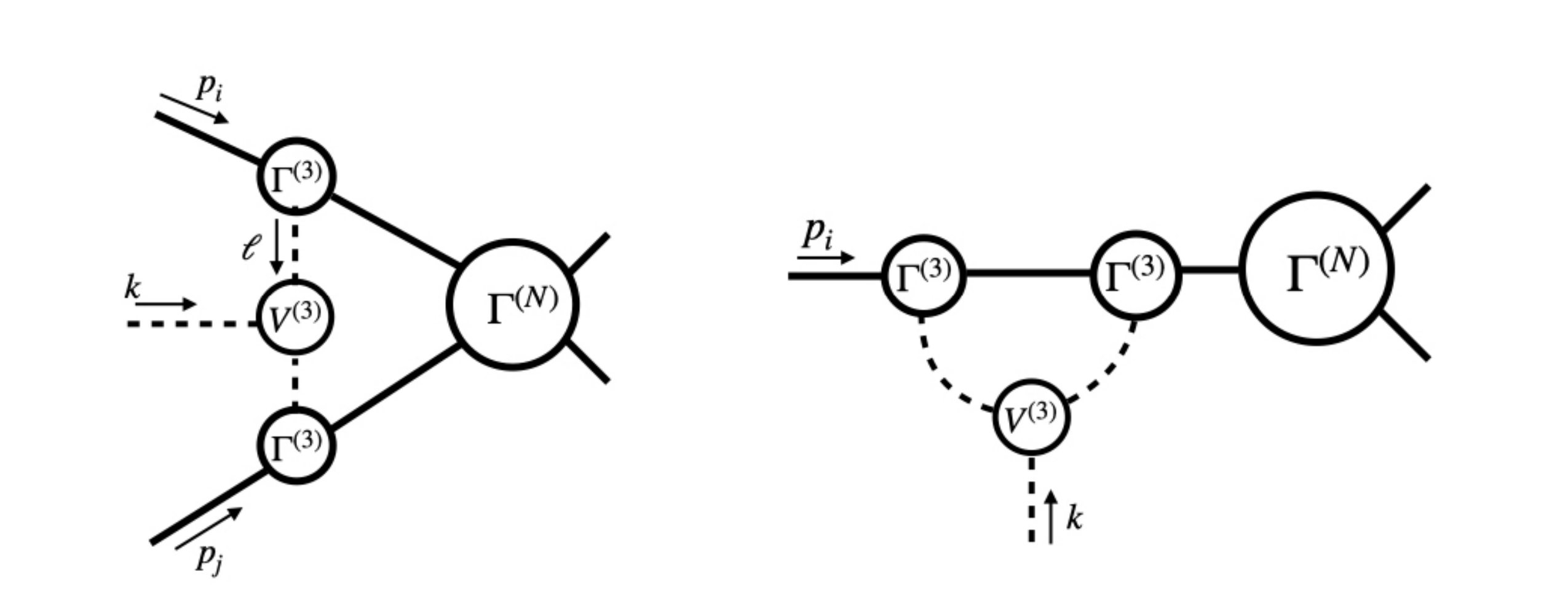}
	\caption{ Diagrams containing 3-gluon vertex contributing to $\mathcal{A}^{(N+1)}$ after regulating the IR-divergence considering detector resolution as the explicit IR cut-off. The solid lines represent massive particles and the dashed lines represent gluons. The picture is taken from \cite{Krishna:2023fxg}.}\label{f:3-gluons_diagrams}
\end{figure}
\end{center}

  Here we are mostly interested in studying the leading soft theorem at one loop. The relevant question we want to address is the logarithmic correction to the $\frac{1}{\omega}$ behavior. The leading soft theorem will be proportional to $\frac{\mathrm{Log}[\omega]}{\omega}$ where $\omega$ is the energy of the real soft gluon emitted. Ideally one should be able to demonstrate this in QCD with fermions and scalars. But as a proof of principle, we will carry out the calculation in scalar chromodynamics where the scalars are in the fundamental representation of the gauge group. We will simply sketch the logarithmic corrections to the leading soft theorem. The contributions come from three gluon vertices. We will analyze the diagrams drawn in fig.\ref{f:3-gluons_diagrams}. We will introduce the explicit IR cutoff $R^{-1}$ which is like the resolution of the detector in the loop momentum integration. We will split the momentum into three regions as $|\ell^\mu|\in [R^{-1},\omega]$, $|(\ell+k)^\mu|\in [R^{-1},\omega]$, and  ``$reg$''$\equiv |\ell^\mu|\in [\omega, |p_i^\mu|]$. In the soft limit, using the symmetry of the first diagram in Fig. \ref{f:3-gluons_diagrams}, the momentum integration in the regions $|\ell^\mu|\in [R^{-1},\omega]$ contributes equally to $|(\ell+k)^\mu|\in [R^{-1},\omega]$. After simplifying the propagators in the appropriate regions we have the expected $\frac{\mathrm{Log}[\omega]}{\omega}$ contributions
\begin{eqnarray}
 &\propto&g_{YM}^2 \sum_{\substack{i,j=1\\i\neq  j}}^{N}\frac{1}{p_j.k}\int_{R^{-1}}^\omega \frac{d^4\ell}{(2\pi)^4}\  \frac{1}{\ell^2-i\epsilon}\frac{1}{p_i.\ell+i\epsilon}\frac{1}{\ell.k-i\epsilon}\Big[ (p_j.k) (p_i.\varepsilon)-2(p_i.k)(p_j.\varepsilon)\Big]\ \Gamma^{(N)}_{(ij)}(p_i,p_j)\nonumber\\
 \label{three gluon}   
\end{eqnarray}
The integrals can be evaluated following \cite{Sahoo:2018lxl}.
\begin{eqnarray}
 &&\int_{R^{-1}}^{\omega} \frac{d^4\ell}{(2\pi)^4}\  \frac{1}{\ell^2-i\epsilon}\frac{1}{p_i.\ell+i\epsilon}\frac{1}{\ell.k-i\epsilon}\nonumber\\
&=& -\frac{1}{4\pi}\frac{1}{p_i.k}\ln(\omega R)\left[\delta_{\eta_i,-1}\ -\frac{i}{2\pi}\ \ln\left(\frac{p_i^2}{(p_i.\mathbf{n})^2}\right)\right]+\mathcal{O}(\omega^{-1})\ .   
\end{eqnarray}

After doing the integrals, we have 
\begin{eqnarray}
 &\propto&\frac{-g_{YM}^2}{4 \pi}\mathrm{log}[\omega R] f_{abc}\sum_{\substack{i,j=1\\i\neq  j}}^{N}T^b_iT^c_j\Bigg(\frac{p_i.\varepsilon}{p_i.k}-\frac{ p_j.\varepsilon}{p_j.k}\Bigg)\left[\delta_{\eta_i,-1}\ -\frac{i}{2\pi}\ \ln\left(\frac{p_i^2}{(p_i.\mathbf{n})^2}\right)\right]\ \Gamma^{(N)}_{(ij)}(p_i,p_j)\nonumber\\   
\end{eqnarray}
the convention here is $\eta_j = +1$ for the ingoing and $\eta_j = -1$ for outgoing particles.\\

There is an easy and ad hoc way to find these factors directly from the results obtained from dimensional regularization prescriptions. We will start with the soft current obtained in eq. 53 of \cite{Catani:2000pi}. The finite contribution can be extracted from expanding $\Bigg(\frac{4 p_i.p_j}{2 p_i.k \, p_j.k\,  e^{- i \pi}}\Bigg)^{\epsilon}$. The finite part of the soft current upon contraction with polarization can be written as
\begin{eqnarray}
    \varepsilon\cdot J_a=-\frac{1}{16 \pi^2}i f_{abc} \sum_{i\neq j} T_i^b T_j^c \varepsilon_{\mu} \Bigg(\frac{p_i^{\mu}}{p_i.k}-\frac{p_j^{\mu}}{p_j.k}\Bigg)\mathrm{Log}^2\Bigg(\frac{4 p_i.p_j}{2 p_i.k\,  p_j.k\, e^{- i \pi}}\Bigg)
\end{eqnarray}
 But in Grammer-Yennie formalism the soft factor comes from the $G$ gluon part. The $K$ gluon part which exponentiates also has a finite term which goes as $\mathrm{Log}[\omega]$. The easiest way to see this is to look at the explicit form of $ \mathrm{K_{glu}}$.
\begin{eqnarray}
 \mathrm{K_{glu}}=  \ \f{i}{2}\sum_{i=1}^{N}\  \sum_{\substack{j=1\\j\neq i}}^{N}T_{i}T_{j}\ \int \f{d^{4}\ell}{(2\pi)^{4}} \ \f{1}{\ell^{2}-i\epsilon}\ {(2p_i-\ell)\cdot(2p_j+\ell) \over (2p_i\cdot\ell -\ell^2+i\epsilon) 
(2p_j\cdot \ell+\ell^2-i\epsilon)}   \end{eqnarray}
The integral runs from $|\ell| \in [0,\infty]$. But we can break the regions of integrations like $|\ell^\mu|\in [0,\omega]$ and  ``$reg$''$\equiv |\ell^\mu|\in [\omega, |p_i^\mu|]$. The regions  $|\ell^{\mu}|\in [0,\omega]$ have IR divergences that we have already taken care of by taking the finite part of the dimensionally regularized integral. The other regions ``$reg$''$\equiv |\ell^\mu|\in [\omega, |p_i^\mu|]$ will give us $\mathrm{Log}[\omega]$ as finite contributions. It contains all the other parts as well. But in our ad-hoc approach we will just take the $\mathrm{Log}[\omega]$ as finite contributions. Hence, the finite soft factor from dimensionally regularized theory will look like 
\begin{eqnarray}
    \varepsilon\cdot J_a&&\propto-\frac{1}{16 \pi^2}i f_{abc} \sum_{i\neq j} T_i^b T_j^c \varepsilon_{\mu} \Bigg(\frac{p_i^{\mu}}{p_i.k}-\frac{p_j^{\mu}}{p_j.k}\Bigg)\mathrm{Log}\Bigg(\frac{4 p_i.p_j}{2 p_i.k\,  p_j.k\, e^{- i \pi}}\Bigg)\mathrm{Log}\Bigg(\frac{4 p_i.p_j}{2 p_i.k\,  p_j.k\, e^{- i \pi}}\Bigg)\non\\
    &&\propto -\frac{1}{16 \pi^2}i f_{abc} \sum_{i\neq j} T_i^b T_j^c \varepsilon_{\mu} \Bigg(\frac{p_i^{\mu}}{p_i.k}-\frac{p_j^{\mu}}{p_j.k}\Bigg)\mathrm{Log}\Bigg(\frac{4 p_i.p_j}{2 p_i.k\,  p_j.k\, e^{- i \pi}}\Bigg)\mathrm{K_{glu}}\non\\
\end{eqnarray}
In the Grammer-Yennie prescription for soft theorem, we factor out the $\mathrm{K_{glu}}$. Hence the soft factor has only $\frac{\mathrm{Log}[\omega]}{\omega}$ dependence. This is not the best way to find out the soft factor. Earlier in this section, we sketched how one can do explicit calculations to find out these soft factors. This is exactly analogous to the case for soft factor in QED and gravity which is already discussed in \cite{Sahoo:2018lxl,Krishna:2023fxg,Sahoo:2020ryf}. \\

\section{Some useful integrals}
\label{identities}
Here we have written some of the integrals and identities that we have used for the mode algebra of graviton modes. 
 \begin{eqnarray}
      \label{identity z1b contour}
      \oint_{|\bar{z}_1|< \epsilon} \frac{d \bar{z}_1}{2 \pi i} \bar{z}^{n+\frac{k-3}{2}}_1 \bar{z}_{12}^m&&= 0 ,\quad 0\leq m <\frac{1-k}{2}-n\non\\
    &&= \frac{m!}{(\frac{1-k}{2}-n)!(m+n+\frac{k-1}{2})!}(-\bar{z}_2)^{m+n+\frac{k-1}{2}}, \quad \frac{1-k}{2}-n \leq m\leq  1-k\non\\
 \end{eqnarray}

The other identity we will use is the contour integral with $\bar{z}_2$. 
\begin{eqnarray}
  \label{contour z2b without log}
&&\oint_{|\bar z_2|< \epsilon} \frac{d \bar{z}_2}{2\pi i}\bar z_2^{m+n+ \frac{k-1}{2}+n'+\frac{l-3}{2}} \partial^m_{\bar z_{2}} R^{k+l-1, c}(z_2, \bar z_2) \non\\
 && = \oint_{|\bar z_2|<\epsilon}  \frac{ d \bar z_2}{2 \pi i}\bar z_2^{m+\frac{k-1}{2} +n+n'+ \frac{l-3}{2}} \partial^m_{\bar z_2} \sum_{m'={\frac{k+l-2}  {2}}}^{\frac{2-k-l}{2}}\frac{R^{k+l-1,c}_{m'}(z_2)}{\bar z_2^{m'+\frac{k+l-2}{2}}}\non\\ 
&& = \frac{(\frac{1-k}{2}- n+ \frac{1-l}{2}-n')!}{(\frac{1-k}{2}- n+ \frac{1-l}{2}-n'-m)!}  R^{k+l-1,c}_{ n+n' }(z_2).
\end{eqnarray}

Some of the identities we have used when summing over the descendants are 
\begin{eqnarray}
\label{2F1}
{}_2F_1(a,b;c;1)=\frac{\Gamma(c) \Gamma(c-a-b)}{\Gamma(c-1)\Gamma(c-b)}, \qquad Re(c)> Re(a+b)
\end{eqnarray}
The identities of the Gamma function which we have used 
\begin{eqnarray}
    \Gamma(z-n)=(-1)^n \frac{\Gamma(-z)\Gamma(1+z)}{\Gamma(n+1-z)}, \quad n \in \mathbb{Z}
\end{eqnarray}

 \bibliographystyle{JHEP}
\bibliography{ref}

\end{document}